\documentclass[runningheads]{llncs}
\usepackage{graphicx}
\usepackage{amsmath}
\usepackage{multirow}
\usepackage{makecell}
\usepackage{amssymb}
\usepackage{marvosym}
\usepackage{booktabs}
\usepackage{cite}
\usepackage{color}
\usepackage{verbatim}
\usepackage{threeparttable}
\usepackage[marginal]{footmisc}

\begin{document}

\titlerunning{nnU-Net and radiomics for segmentation and prognosis}

\title{Joint nnU-Net and Radiomics Approaches for Segmentation and Prognosis of Head and Neck Cancers with PET/CT images }
%
%
\author{Hui Xu\inst{1,2,*}
\and
Yihao Li\inst{2,*}
\and
Wei Zhao\inst{3}
\and
Gwenolé Quellec\inst{2}
\and
Lijun Lu\inst{1}\textsuperscript{(\Letter)}
\and
Mathieu Hatt\inst{2}
}
\authorrunning{H. Xu et al.}
%
\institute{School of Biomedical Engineering, Southern Medical University, Guangzhou, China \\
\email{ljlubme@gmail.com}\\
\and
LaTIM, INSERM, UMR 1101, Univ Brest, Brest, France
\and
Department of Colorectal Surgery, Chinese Academy of Medical Sciences and Peking Union Medical College, Beijing, China\\
}
\maketitle              
\footnote{Team name: RokieLab. \\Hui Xu and Yihao Li contribute equally to this work.\\}
\begin{abstract}
Automatic segmentation of head and neck cancer (HNC) tumors and lymph nodes plays a crucial role in the optimization treatment strategy and prognosis analysis. This study aims to employ nnU-Net for automatic segmentation and radiomics for recurrence-free survival (RFS) prediction using pretreatment PET/CT images in multi-center HNC cohort. A multi-center HNC dataset with 883 patients (524 patients for training, 359 for testing) was provided in HECKTOR 2022. A bounding box of the extended oropharyngeal region was retrieved for each patient with fixed size of 224 × 224 × 224 $mm^{3}$. Then 3D nnU-Net architecture was adopted to automatic segmentation of primary tumor and lymph nodes synchronously.Based on predicted segmentation, ten conventional features and 346 standardized radiomics features were extracted for each patient. Three prognostic models were constructed containing conventional and radiomics features alone, and their combinations by multivariate CoxPH modelling. The statistical harmonization method, ComBat, was explored towards reducing multicenter variations. Dice score and C-index were used as evaluation metrics for segmentation and prognosis task, respectively. For segmentation task, we achieved mean dice score around 0.701 for primary tumor and lymph nodes by 3D nnU-Net. For prognostic task, conventional and radiomics models obtained the C-index of 0.658 and 0.645 in the test set, respectively, while the combined model did not improve the prognostic performance with the C-index of 0.648. 

\end{abstract}
\section{Introduction}
Head and Neck cancer (HNC) is the fifth most common cancer worldwide \cite{parkin2005global}. Radiotherapy combined with cetuximab has been established as standard treatment \cite{bonner2010radiotherapy}; however, 40\% of patients still experience loco-regional failures in the first two years after the treatment \cite{chajon2013salivary}. Early prediction of prognosis response is crucial to tailor individualized treatment strategies for improving long-term survival of HNC patients. Positron Emission Tomography/Computed Tomography (PET/CT) imaging were reported as a powerful tool in managing HNC patients including diagnosis, staging, design of the radiotherapy planning, and prognosis evaluation \cite{goel2017clinical}. Besides conventional metrics (i.e., TNM stage, tumor volume, SUV), radiomics features were also considered with predictive values in clinical decision-making for HNC patients. For instance, Vallières  et al. demonstrated the potential of radiomics for assessing the risk of specific tumour outcomes using multiple stratification groups \cite{vallieres2017radiomics}. Bogowicz et al. investigated the prognostic value of both PET and CT radiomics features, and they indicated their combinations showed better discriminative power than individual modality alone for local tumor control modelling in HNC\cite{bogowicz2017comparison}. Despite with promising benefits, both radiotherapy, conventional features and radiomics approaches are heavily dependent on the accurate identification and segmentation of the primary tumor and lymph node regions, which are mostly utilized manual annotations in previous studies. However, this process is extremely time consuming and suffers from inter-observer variability. Thus, an automatic segmentation method \cite{oreiller2022head} would greatly assist in formulating radiotherapy plans and early assessing prognosis via quantitative metrics, such as radiomics or other advance techniques. \\

The HEad and neCK TumOR (HECKTOR) challenge 2021 was organized within the context of MICCAI 2021 and provided opportunity for participants to segmentation of primary tumor and prognosis prediction \cite{andrearczyk2021overview}. However, for a convincing validation and promote the transformation of clinical application, both segmentation and prognosis models need to be validated on larger and multi-center cohorts. Moreover, expansion of radiomics analysis with lymph nodes features was reported with significant improvement in prognosis prediction compared to the analysis of primary tumor alone \cite{bogowicz2019combined}. In this context, the challenge HECKTOR 2022 contains more data and extends the scope of investigation \cite{andrearczyk2022overview}. First, lymph nodes were added into the segmentation task. Second, a larger multi-center cohort was provided with 883 HNC patients totally. Third, no bounding boxes are provided to pursue a fully automatic pipeline. The challenge of HECKTOR 2022 mainly contains two tasks, which are 1) the automatic segmentation of primary tumor Gross Tumor Volumes (GTVp) and lymph nodes (GTVn) (Task 1: segmentation task), and 2) early prediction of Recurrence-Free Survival (RFS) for HNC patients (Task 2: prognosis task).\\

\begin{figure}[!h]
\centering
\includegraphics[width=\textwidth]{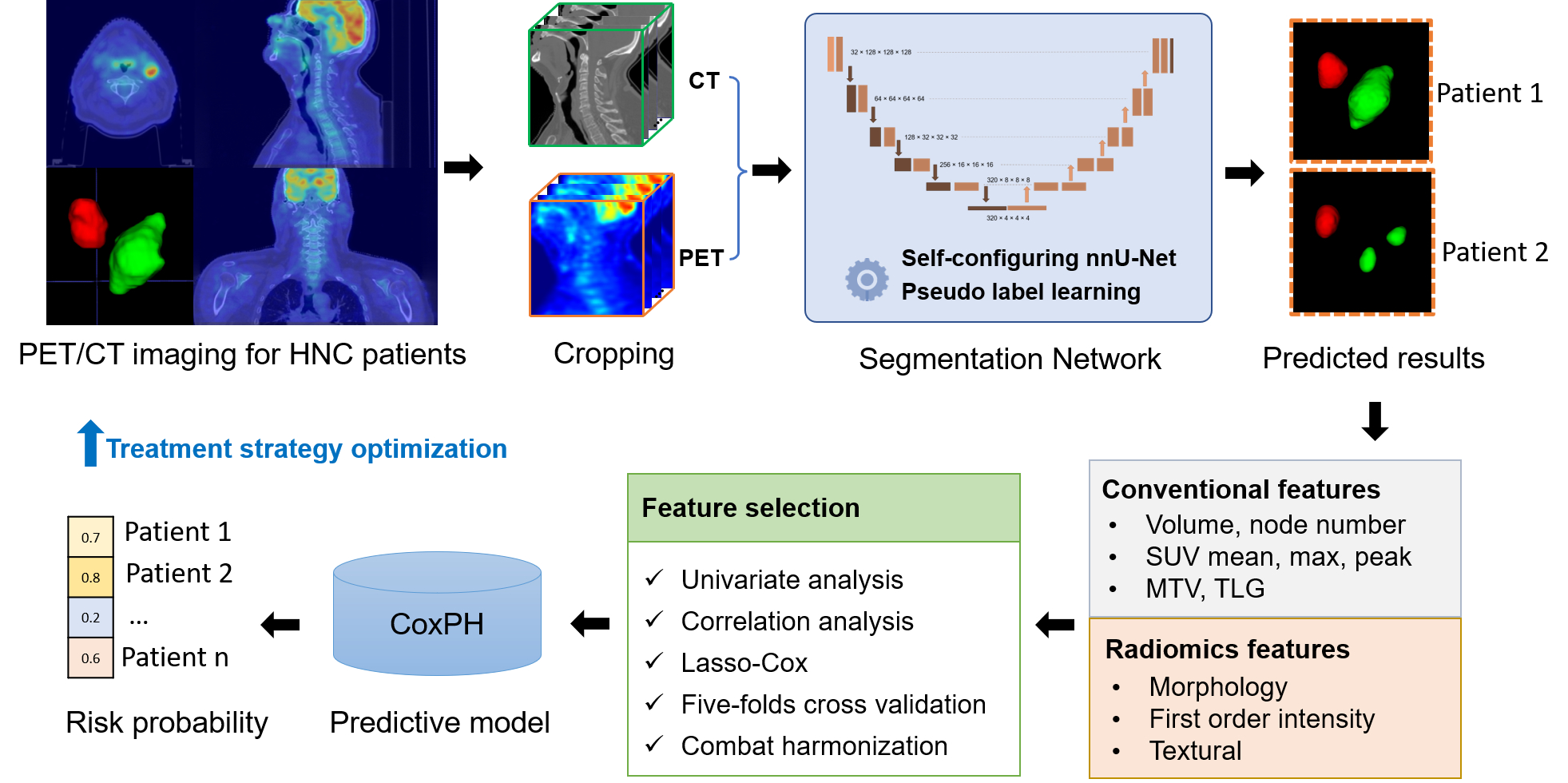}
\caption{The workflow of this study including images pre-processing, segmentation and prognosis prediction.} \label{fig1}
\end{figure}

In this study, we adopt 3D nnU-Net as our baseline model for segmentation task since it was demonstrated with excellent results in biomedical images segmentation without manual intervention \cite{isensee2021nnu}, and has shown good performance in HECKTOR 2021 \cite{murugesan2021head}. Additionally, we propose a modified version using pseudo labelling technique followed by 3D nnU-Net. Based on the predicted segmentations of GTVp and GTVn, we construct Cox Proportional Hazards regression model (CoxPH) combined with conventional and/or radiomics features for predicting the risk of RFS in a multi-center HNC cohort. The workflow of this study is shown in Fig.~\ref{fig1}.

\section{Method}

\subsection{Dataset description}

This dataset consisted of 883 patients from 9 centers with histologically proven oropharyngeal HNC cancer. A total of 524 patients from 7 centers was used as training set for Task 1. Of these, 489 patients were used Task 2. The test set contained 359 patients from 3 centers that are unseen in previous year’s challenge \cite{andrearczyk2021overview}; specifically, 359 and 339 patients were used for Task 1 and Task 2, respectively. Of note, the center MDA is represented both in the training and test sets. Details of dataset are provided in Table~\ref{tab1}. In the training set, pre-treatment FDG PET/CT images, segmentation mask including primary tumor (GTVp) and lymph nodes (GTVn), clinical and prognosis information were provided for each patient. All the image dataset is NIfTI format, and the segmentation mask has the same resolution with CT images. Clinical data including center, gender, age, weight without missing was provided. In the test set, only pre-treatment PET/CT images and clinical information were provided.

\begin{table}
\centering
\caption{The statistic and partitioning of multi-center HNC datasets.}\label{tab1}
\begin{tabular}{cccccc}
\toprule  
No. & Center & Devices & Task 1 & Task 2  & Cohort \\
\midrule  
1 &  CHUM  &Discovery STE, GE Healthcare &56 & 56 & Training\\
2 &  CHUP & Biography mCT 40 ToF, Siemens& 72& 44 & Training\\
3 & CHUS  & Gemini GXL 16, Philips & 72& 72 & Training\\
4 &  CHUV & Discovery D690 ToF, GE Healthcare& 53 & 47 & Training\\
5 &  MDA & \makecell[c]{Discovery HR, Discovery RX, Discovery ST,\\ Discovery STE (GE, Healthcare)} & 198/200 & 197/200 & Training/Test\\
6 &  HGJ & Discovery ST, GE Healthcare & 55 & 55 & Training\\
7 &  HMR & Discovery STE, GE Healthcare & 18 & 18 & Training\\
8 &  USZ & \makecell[c]{Discovery HR, Discovery RX, Discovery STE, \\ Discovery LS, Discovery 690 (GE Healthcare)} & 101 & 101 & Test\\
9 &  CHB & GE710, GE Healthcare & 58 & 58 & Test\\
\bottomrule 
\end{tabular}
\end{table}

\subsection{Pre-processing}

Since the covered regions of images are not consistent between patients, we adopted an automatic method \cite{andrearczyk2020oropharynx} to retrieve head and neck regions. This method relies on anatomical information and PET intensity as prior to find brain region, In our study, a fixed size bounding box of $224 \times 224 \times 224 \; mm^3$ located three centimeters shift downward and forward from the lowest voxel of brain region was first determined. We evaluated the results by checking whether the GTVp and GTVn are fully contained by the bounding box according to the provided segmentation, where 509 out of 524 (97.1\%) cases were correctly detected. In testing set, we only checked that the intensity range of PET and CT images is normal and not full zero. Here 353 out of 356 (99.2\%) cases were correctly detected. For these failure cases, a semi-automatic method by setting the center voxel by our experience was adopted to get the location of bounding box with same size. Once the bounding box location was confirmed, we cropped and resampled the original PET, CT and mask images to same scale of $224 \times 224 \times 224$ with the isotropic voxel size of $1 \times 1\times 1 \;mm^3$ by linear interpolation. No further intensity normalization was performed on CT and PET images.  

\subsection{Task 1: Segmentation Prediction}

\subsubsection{Network Architecture} 

\begin{figure}[!t]
\centering
\includegraphics[width=\textwidth]{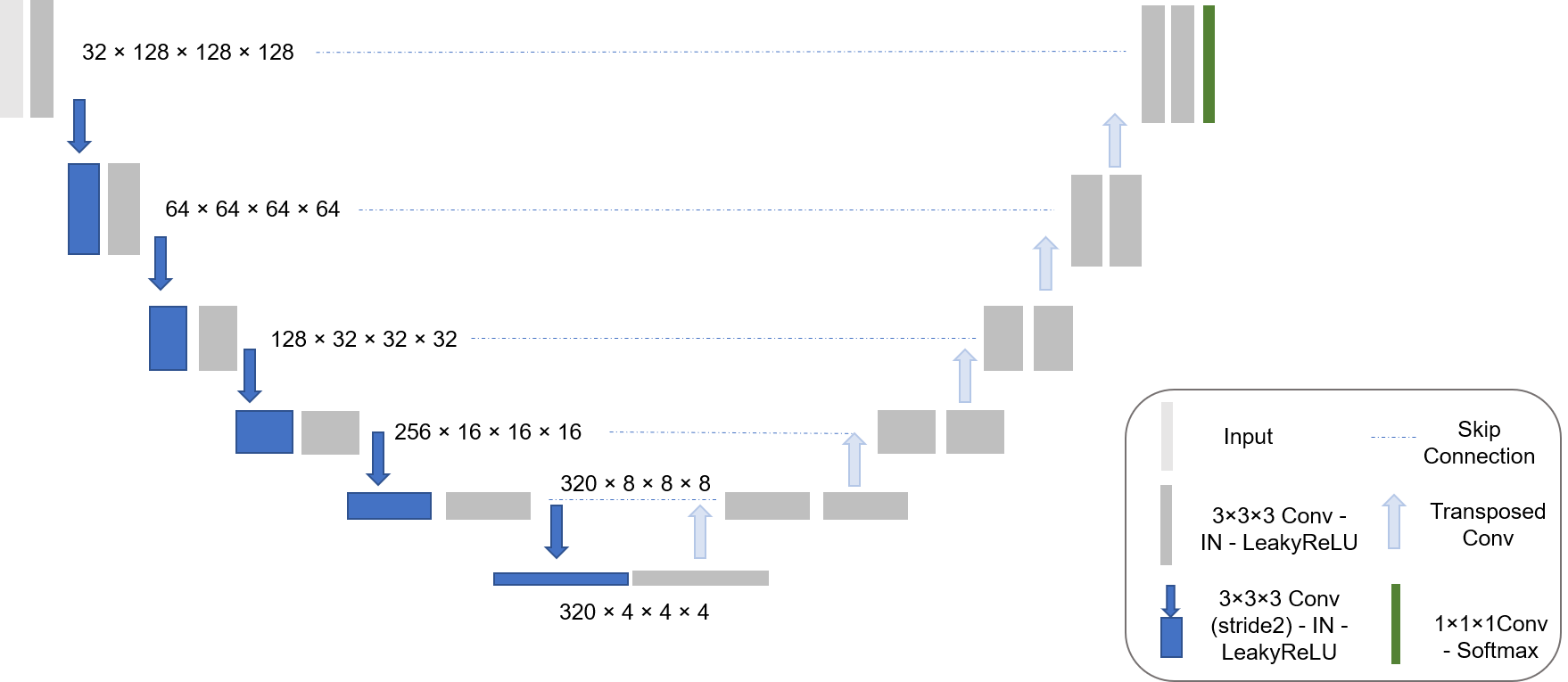}
\caption{The architecture of the 3D nnU-Net.} \label{fig2}
\end{figure}

The architecture of the 3D nnU-Net is shown in Fig.~\ref{fig2}. Before the launch of segmentation process, nnU-Net cropped the non-zero area of the PET and CT bounding box automatically. Then a patch-based sliding window technique was applied to the current cropped images, producing the patches with size of $128 \times 128 \times 128$. These patches were input into the 3D nnU-Net. Two Conv-InstanceNorm-LeakyReLU blocks for down-sampling and up-sampling of the encoder and decoder was used in nnU-Net. Down-sampling is done by strided convolution, while up-sampling was done by transposed convolution. The architecture initially used 30 feature maps, which are doubled for each down-sampling operation in the encoder (up to 320 feature maps) and halved for each transpose convolution in the decoder. The end of the decoder has the same size as the input, followed by a $ 1 \times 1 \times 1$ convolution and a soft-max function.

\subsubsection{Implementation Details} 
A combination of dice and cross-entropy loss is used to train our networks:

\begin{equation}
L_{total} = L_{dice} + L_{CE}  
\end{equation}

The dice loss formulation used here is adapted from the variant proposed in \cite{https://doi.org/10.48550/arxiv.1809.10486}. It is implemented as follows:

\begin{equation}
L_{dice} = - \frac{2}{\left | K \right | } \sum_{k\in K}^{} \frac{ {\textstyle \sum_{i\in I}^{}} u_{i}^{k}v_{i}^{k}  }{ {\textstyle \sum_{i\in I}^{}} u_{i}^{k}+  {\textstyle \sum_{i\in I}^{}} v_{i}^{k}   } 
\end{equation}

where $u$ is the soft-max output of the network and $v$ is the one hot encoding for the ground truth segmentation map. Both $u$ and $v$ have shape $I \times K $  with $i \in I$ being the pixels in the training patch/batch and $k \in K$ being the classes.\\

Adam optimizer was used with an initial learning rate of $ 3 \times 10^{-4}$. The epoch is 1000 with one epoch being defined as iteration over 250 minibatches. Batch size is 2. A five-fold cross-validation procedure was used, and 1000 epochs were trained per fold. Whenever the exponential moving average of the training losses did not improve by at least $ 5 \times 10^{-3}$ within the last 30 epochs, the learning rate was reduced by factor 5. The training was terminated automatically if the exponential moving average of the validation losses did not improve by more than $ 5 \times 10^{-3}$ within the last 60 epochs, but not before the learning rate was smaller than $ 10^{-6}$ \cite{https://doi.org/10.48550/arxiv.1809.10486}.\\

During training, the following augmentation techniques were applied: random rotations, random scaling, random elastic deformations, gamma correction augmentation and mirroring. NVIDIA RTX A6000 48G was used in training. During inference, similar sliding window method was first used to generate patches: four $128 \times 128 \times 128$ voxel patches were processed and the predicted output probabilities were averaged in the overlapping regions. The training took about 20 hours per fold, and the inference on the test set took approximately 30 minutes.\\

\subsection{Task 2: Prognosis Prediction}

\subsubsection{Conventional features development} 

Previous studies have reported the clinical prognostic values of conventional features in PET/CT imaging \cite{picchio2014predictive}, thus a total of ten conventional features were firstly calculated using an in-house developed package based on the predicted segmentations of 3D nnU-Net. Conventional features included primary tumor volume, diameter, number of nodes, maximum, mean and peak standardized uptake value (SUVmax, SUVmean, SUVpeak), metabolic tumour volume based on the threshold of SUV2.5 and 40\% SUVmax separately (MTV2.5, MTV40\%), and total lesion glycolysis using corresponding MTV multiply by SUVmean (TLG2.5, TLG40\%). Specifically, these parameters related to SUV were calculated across both primary tumor and lymph nodes regions.

\subsubsection{Radiomics features extraction} 

Radiomics features were extracted from both PET and CT images separately using the open-source package of Standardized Environment for Radiomics Analysis (SERA) (\url{https://github.com/ashrafinia/\\SERA}), which conform to the image biomarker standardisation initiative (IBSI) \cite{zwanenburg1612image}. Based on predicted segmentation using 3D nnU-Net, all default features were extracted from the region including primary tumor and lymph nodes with setting of isotropic voxel sizes of 2 × 2 × 2 $mm^{3}$ in resampling and fixed bin number of 64 in discretization, which resulted in totally 346 radiomics features for each patient. 

\subsubsection{Cross-Validation Strategy} 

To evaluate the generalization performance of our prognostic models, we created patient folds from the training set as validation set by using cross validation strategy. This step was used for feature selection and hyper-parameters adjustment in modelling. Since the provided testing set contained 200 patients from the center MDA, we randomly chose 97 patients ($\sim$20\% of all training data) from MDA center in the training set used as a separate fold namely fold 5. This set aimed to select the prognostic model (developed by the training set) with relatively consistent performance in testing set. The remained training set (392 patients) were randomly split into four folds, namely fold 1, 2, 3, 4 (98 patients for each fold). The mean performance of five folds helps us to pick up prognosis models for submission.

\subsubsection{Model construction}

Three prognostic models were submitted which are based on 1) conventional features alone; 2) radiomics features alone combined with Combat harmonization strategy \cite{johnson2007adjusting}; 3) conventional features combined with radiomics features without harmonization. For model 1, feature selection included univariate CoxPH model keeping features with the concordance index (C-index)\cite{harrell1996multivariable} higher than 0.50, and Pearson correlation analysis to remove redundant features ($\rho_{c} <$ 0.60). For model 2, Combat harmonization of non-parametric version  \cite{johnson2007adjusting} was firstly applied for all radiomics features within the joint training and test sets using their center labels. Patient’s gender, age and weight were used in the design matric to depict biological covariate(s) in Combat harmonization. Followed by the similar process as conventional features in feature selection including univariate and correlation analysis, the least absolute shrinkage and selection operator Cox regression algorithm (Lasso-Cox) was adopted to identify the optimal features set. For model 3, the consistent features selection was used for radiomics features as model 2, while these features have not been applied with Combat harmonization. Then the optimal radiomics and conventional features set were integrated as prognostic predictors. After feature selection, multivariate CoxPH model was constructed based on all training set to predict the survival risk (RFS) of HNC patients.

\subsection{Evaluation Metrics}

Aggregated Dice score \cite{kumar2017dataset} from both GTVp and GTVn was used as evaluation metric for segmentation task. The concordance index (C-index) between the predicted risk and the survival outcomes was used to measure the predictive performance in prognosis task.

\begin{table}
\centering
\caption{Segmentation results (Dice score) of two submitted models in the training and test sets.}\label{tab2}
\setlength{\tabcolsep}{2.25mm}{
\begin{tabular}{|c|c|c|c|c|c|c|}

\cline{1-7}  
\multirow{2}*{Model} & \multicolumn{3}{c}{Training} & \multicolumn{3}{c|}{Testing}  \\
\cline{2-7}
 &  GTVp  &GTVn &mean & GTVp & GTVn&mean\\
\hline
nnU-Net (baseline) &  0.86724 &0.81576 &0.84150 & 0.69997 & 0.70039&0.70018\\
\hline
nnU-Net with PLL &  0.82551 & 0.75506& 0.79028 & 0.70131&0.70100&0.70115\\
\hline
\end{tabular}}
\end{table}

\section{Results}

\subsection{Task 1: Segmentation results}

\begin{figure}[!h]
\centering
\includegraphics[width=\textwidth]{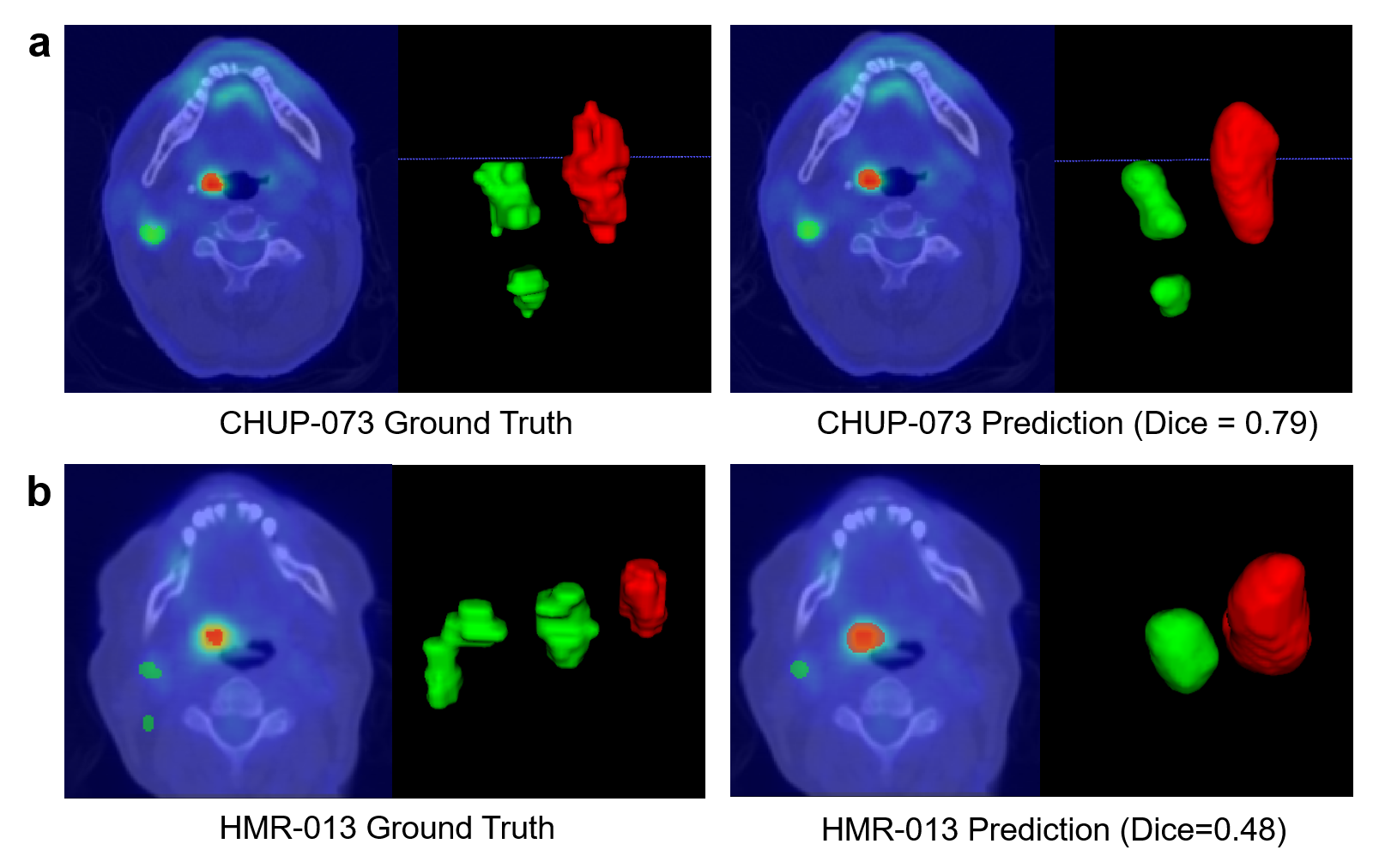}
\caption{An example of two patients (CHUP-073, HMR-013) for comparing the ground truth segmentation by manual and the predicted segmentation by 3D nnU-Net.} \label{fig3}
\end{figure}

Table~\ref{tab2} provides the segmentation results (Dice score) of GTVp, GTVn and their averaged value on the training and test sets. The baseline segmentation model using 3D nnU-Net showed a mean Dice score of 0.70018 across primary tumor and lymph nodes regions. Once baseline nnU-Net model was trained, We further adopted the pseudo-labeling learning (PPL) technique to enhance the performance of segmentation model. PPL technique consists in adding confidently predicted test data to the training data to optimize parameters learning \cite{lee2013pseudo}. Thus, 3D nnU-Net with PPL was combined to retrain nnU-Net model which integrated original training set and the predicted testing set as an updated training set. The updated model led to the mean dice of 0.70115 in the test set. Overfitting is caused mainly by the fact that the data center for the test set and training set is different. An example with good and poor segmentation quality separately by baseline nnU-Net is displayed in Fig.~\ref{fig3}.

\begin{table}
\centering
\caption{The individual performance of ten conventional features in five-folds cross validation.}\label{tab3}
\setlength{\tabcolsep}{2mm}{
\begin{tabular}{|c|c|c|c|c|c|}
\hline
 Features &  Volume  & Diameter & \makecell[c]{Number of \\ nodes} & SUVmax & SUVmean\\
\hline
Mean C-index &  0.5736 &0.5611 &0.6029 & 0.5924 & 0.5926\\
\hline
Features &  SUVpeak & MTV2.5& MTV40\% & TLG2.5&TLG40\%\\
\hline
Mean C-index &  0.6028 & 0.6225& 0.6211 & 0.6393&0.6295\\
\hline
\end{tabular}}
\end{table}

\subsection{Task 2: Prognostic prediction results}

The individual performance of 10 conventional features in the training set by five folds cross validation is listed in Table~\ref{tab3}. Their performances in testing set were not verified due to the limited availability of the testing set. After feature selection, two conventional features (number of nodes, TLG2.5) were remained to build prognostic model (Model 1), which showed the C-index of 0.6582 in independent testing. The overall results of our submitted three models is provided in Table~\ref{tab4}. For radiomics features with Combat harmonization, eight features (listed in Table 4) were combined to predict prognosis (Model 2) with the C-index of 0.6452 in testing. For radiomics features without harmonization, only two features were selected by the designed feature selection strategy. We further combined them with two conventional features to build prognostic model (Model 3), which showed the improved performance in the training set but slightly decreased performance in testing (C-index, 0.6475).

\begin{table}
\centering
\caption{The performance of conventional, radiomics and combined models for RFS prediction.}\label{tab4}
\begin{threeparttable}
\begin{tabular}{|c|c|c|c|c|}
\hline
 Model &  Features  & \makecell[c]{Validation \\ (5-fold)} & Training & Testing\\
\hline
Conventional &  number of nodes, TLG2.5 &0.6716 &0.6590 & 0.6582 \\
\hline
Radiomics (with Combat) &  \makecell[c]{CT-Morphology-spherical \\ disproportion, \\CT-Local-peak, \\CT-IH-qcod, \\PET-Statistic-skewness, \\CT-GLSZM-szhge, \\PET-GLDZM-zdnu,  \\PET-NGTDM-coarseness, \\PET-NGLDM-lgce} & 0.6453& 0.6777 & 0.6452\\
\hline
\makecell[c]{Conventional+Radiomics\\(without Combat) } & \makecell[c]{number of nodes, TLG2.5,  \\ CT-Local-peak, \\PET-GLCM-correlation1} & 0.6810& 0.6860 & 0.6475 \\
\hline
\end{tabular}

\begin{tablenotes}
\footnotesize
\item[] Aggregation: IH, Intensity Histogram; qcod, quartile coefficient of dispersion; szhge, small zone high grey level emphasis; zdnu, zone distance non-uniformity; lgce, low grey level count emphasis.

\end{tablenotes}
\end{threeparttable}
    
\end{table}

\section{Discussion}

This study delivered a fully automatic and relatively robust approach for segmentation and prognosis prediction based on primary tumor and lymph nodes of HNC patients using pre-treatment PET/CT imaging. Two segmentation models using 3D nnU-Net with and without pseudo labelling learning were trained. The slight improvement was found in model with PLL which achieved the Dice score of 0.70115 in the testing set. Utilizing the predicted segmentation results by baseline nnU-Net, we extracted conventional and radiomics separately and developed three prognostic models via machine learning approach to predict RFS of HNC patients. We obtained the highest C-index of 0.6582 among our submitted models. \\

In order to develop robust segmentation and prognosis models, we mainly focused on the relatively reliable and simple methods, such as the self-configurating nnU-Net and conventional features, which were validated by a vast of previous multi-center studies \cite{https://doi.org/10.48550/arxiv.1809.10486,picchio2014predictive}. Our segmentation task further proves the power of nnU-Net in medical images segmentation, although it did not perform at the top of the rankings in this challenge. \cite{xie2020head} introduced the squeeze \& excitation structure to nnU-Net and successfully improved their performance in last year's challenge. Moreover, attention mechanism was demonstrated with potential in segmentation when combining with convolutional networks \cite{https://doi.org/10.48550/arxiv.2103.10504,https://doi.org/10.48550/arxiv.2102.04306}, which deserves further research. We recognize that our pseudo labelling approach is not perfect despite slight improved performance in testing, since combined original training and predicted testing set will deepen model learning and introduce incorrect information from wrong cases in testing set. It would be ideal to define a threshold value to identify which case can be used for re-training, as follows. For each prediction result, we can make a confidence map, where the value of each pixel is the maximum probability. It can be considered a pseudo-label if no value in the whole confidence map is less than the threshold. This method needs to be further explored and optimized.\\

Although our segmentation of primary tumor and lymph nodes are not very precise compared to manual delineation, our prognostic models based on whatever conventional and radiomics features showed relatively satisfied performance in the multi-center testing set. Our best model employed two conventional features which are number of nodes and TLG2.5 (MTV2.5 × SUVmean) with relatively high stability and generalization. The individual performance of each conventional features in the training set (Table~\ref{tab3}) emphasized the importance of lymph nodes in prognosis prediction including its volume, number, and SUV metrics. This point is consistent with previous research \cite{bogowicz2019combined,kubicek2010fdg}. We did not perform harmonization for conventional features since they were demonstrated with high stability by previous analysis \cite{leijenaar2013stability}. On the other hand, radiomics features after Combat harmonization also exhibit promising results with the C-index of 0.6452 in the testing set. An interesting finding is that the testing result is very closed to the mean performance of five folds cross validation (Table~\ref{tab4}), which may potentially indicate the reduction of multi-center variations and the improvement of model generalization after Combat harmonization. Given the limitation of submitted models, we cannot provide an explicit comparison to demonstrate the effect of Combat harmonization. The further study aimed to explore model performance with and without harmonization need to be performed in further work. Radiomics features without harmonization combined with two conventional features above did not improve the performance in testing that may be explained by the obstacle of high sensitivity of radiomics features. Thus, compared to conventional features, radiomics features still exhibit some flaws such as its low stability although they can depict more diverse and detailed information of malignant tumors. Of note, our radiomics features were extracted from the regions including primary tumor and lymph nodes, rather than from primary tumor and lymph nodes separately, which will introduce different descriptions of tumor heterogeneity that need to be further explored.\\

\section{Conclusion}

In conclusion, our approach provides an automatic, fast and relatively consistent solution for primary tumor and lymph nodes segmentation in HNC patients, and shows potentials to be generally applied for prognosis evaluation by adopting both conventional and radiomics features.

%
%
%

\bibliographystyle{splncs04}
\bibliography{paper7331}
\end{document}